\documentstyle[11pt,newpasp_charbonneau,twoside,epsf]{article}
\def\edcomment#1{\iffalse\marginpar{\raggedright\sl#1\/}\else\relax\fi}
\reversemarginpar

\begin{document}
\title{The STARE Project:  A Transit Search for Hot Jupiters}
\author{Timothy M. Brown$^{1}$ and David Charbonneau$^{2,1}$}
\affil{$^1$High Altitude Observatory, National Center for Atmospheric 
Research, P. O. Box 3000, Boulder, C0 80307, USA; timbrown@hao.ucar.edu}
\affil{$^2$Harvard-Smithsonian Center for Astrophysics, 60 Garden St., Cambridge, MA 02138, USA; dcharbonneau@cfa.harvard.edu}

\begin{abstract}
The STARE instrument is a small aperture, wide-field, CCD-based
telescope that delivers high cadence time series photometry
on roughly 40~000 stars in a typical field centered on the
galactic plane.  In a two-month observing run on a 
field, we obtain sufficient precision on roughly 4~000 stars
to detect a close-in Jupiter-sized companion in an edge-on orbit.
We also used this instrument to detect the planetary transits across
the Sun-like star HD~209458.  The project is now in its third season, 
and we have acquired a large dataset on several fields.  Given the
frequency of close-in extrasolar planets found by the radial velocity
surveys, and the recent confirmation that at least some of these
are indeed gas giants, the STARE project should be able to detect
roughly a dozen Jupiter-sized planets in its existing dataset.
\end{abstract}

\section{Introduction and Motivation}
Radial velocity surveys of nearby F, G, K and M dwarf stars 
have revealed a class of close-in extrasolar massive planets
that orbit their stars with an orbital separation of 
$a \la 0.1$ AU.  
Prior to the transit results for HD~209458,
the radial velocity method has been the only method by which
we have learned anything about these planets. 
The radial velocity technique measures the period, semi-amplitude, and
eccentricity of the orbit, and by inference the semi-major axis.
It also yields a value for the minimum mass, dependent upon the 
assumed value for the stellar mass, but aside from this it
gives no direct information on the structure of the planet itself.
The search to measure the transit photometrically is motivated
by fact that, for a star for which both the radial velocity
and transits are observed, one can estimate both the mass 
(with negligible error due to $\sin i$) and radius of the planet.
These can then be combined to calculate such
critically interesting quantities as the surface gravity and 
average density of the planet, and thus provide 
constraints on structural models for these low-mass
companions.  
Assuming random orbital alignment for systems with $a = 0.05$ A.U. 
and Sun-like primaries, the chance
of a transiting configuration is roughly 10\%.

\section{The STARE Instrument}
The telescope is a Schmidt camera of focal length 286~mm and f/2.9. 
The modest 10~cm aperture images a 6~degree square field of view
onto a 2034~$\times$~2034 pixel CCD with 15~$\micron$ (11~arcsecond) pixels.
The majority of observations are taken through a red 
(approximately Johnson $R$) filter, with supporting observations
in $B$ and $V$ to determine stellar colors.
The data are analyzed via a pixel-weighted photometry scheme,
with post-photometry linear regression to account for gray and
color-dependent extinction.  Prior to July 1999, the main instrument was 
an Aero-Ektar lens.  The new system has a much more Gaussian-shaped point 
spread function, which increased the precision 
and the throughput by a factor of 2.

\begin{figure}
\plottwo{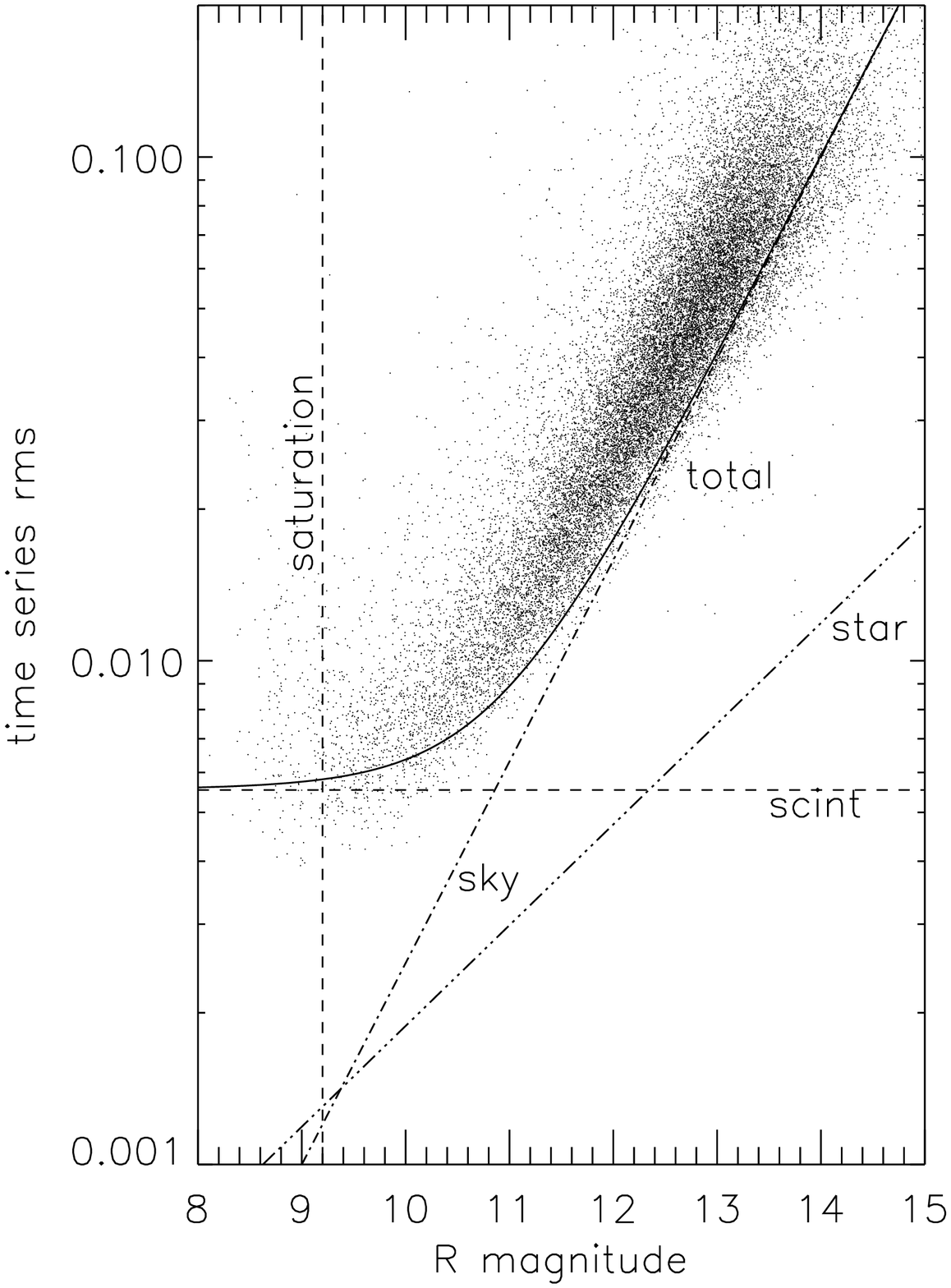}{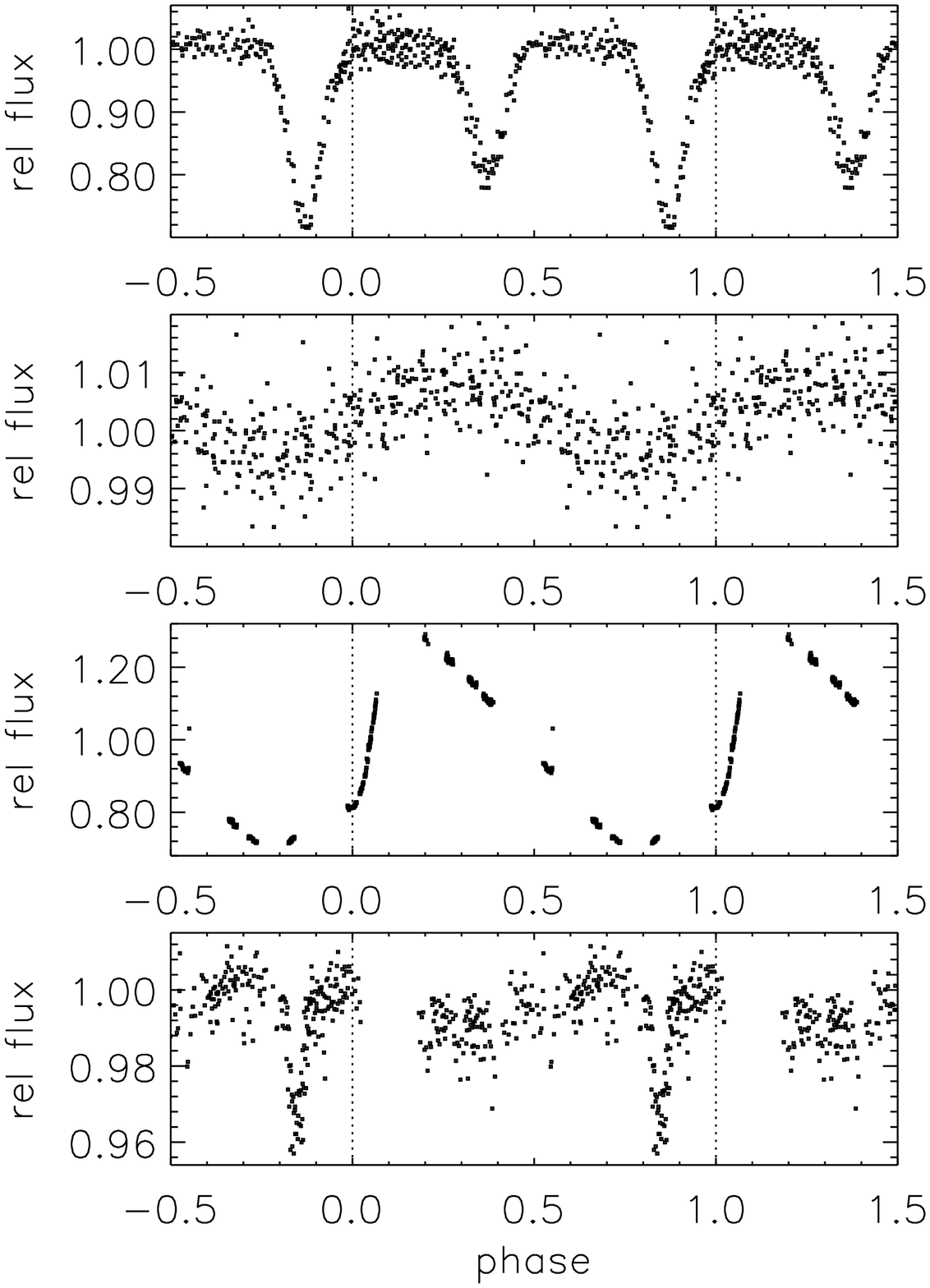}
\caption{The left panel displays the 
achieved time series rms variation for roughly 24~000 stars 
in a field in Auriga, as a function of $R$ magnitude, for one night.  
The dominant noise source is sky background for all but the 
brightest stars in the field.
The right panel shows a 
selection of variable stars as observed by STARE.  From top to
bottom, these are:  1. an eclipsing binary with a period of 0.876 days; 
2. a $\delta$ Scuti with a period of 0.078 days; 3. a $\delta$ 
Cepheid with a period of 18.2 days; 4. an eclipsing 
binary with a 4\% primary eclipse and a period of 2.55 
days.  This small eclipse depth is approaching the level that would
be predicted for a transit of a Jupiter-sized companion
across a Sun-like star.}
\end{figure}

The STARE instrument delivers high cadence (2 minute) times series 
photometry on roughly 40~000 stars (9 $<$ $V$ $<$ 14) in a typical 
field centered on the galactic plane.  As shown in Figure 1, sky
background is the dominant source of noise for the majority
of stars in the field, while the brightest stars are scintillation limited.
In each field, we obtain sufficient precision on roughly 4~000 stars to detect
a close-in Jupiter-sized companion.  Since fields centered on the
galactic plane are not always available, we have also stared at
some high-galactic latitude fields.  These fields are significantly
less crowded and will allow us to measure the effects of crowding
on our photometry pipeline.

The telescope is currently located in Boulder, Colorado, but is
portable.  In the future, it may be relocated in longitude
so as to provide much greater time series coverage when operated
as part of a network with similar instruments, such as those
run by W. Borucki at Lick Observatory 
(Vulcan, Borucki et al. 2000) and by E. Dunham at Lowell Observatory.

A byproduct of the STARE observations is the high-cadence
monitoring of numerous
variable stars, the majority of which are new detections.  We have 
made many of the light curves available electronically (Brown \& Kolinski
1999), and plan to continue to do so in the future.

\section{The First Transiting Extrasolar Giant Planet}
We have detected the first planetary transits across a Sun-like star,
HD~209458, as described in Charbonneau et al. (2000).  
These have also been reported by Henry et al. (2000).

Motivation for observing HD~209458 came from D.~W.~Latham and M.~Mayor
(personal communication) in August~1999.  
The times at which a potential transit could occur were calculated
from the preliminary orbital period and ephemeris from the radial velocity
observations.  We observed HD~209458 for ten nights 
in August and September.  Most of these nights occurred 
when no transit was predicted, and the residuals are consistent
with no variation.  On the other
two nights (UT {9~Sep} \& {16~Sep}), we see a conspicuous dimming of
the star for a time of several hours, at a time consistent
with the prediction from the observed radial velocity orbit.
We attribute this dimming to the
passage of the planet across the stellar disk.  Our precise measurement of
two complete transits allows us to accurately determine the planetary
radius, orbital inclination, and mass, and hence derive quantities such
as the surface gravity and average density.

Assuming a value for the stellar radius and mass, and a description
for the stellar limb-darkening,
we can determine the planetary radius with an accuracy of several percent
(see Figure 2).  However, the dominant uncertainties in determining 
the planetary radius and orbital inclination are the uncertainty 
in the value of the stellar radius, and to a lesser extent the 
stellar mass and limb-darkening.  Given time series observations of 
the transit in a single photometric band, it is possible to fit the data 
with a family of models, since a larger planetary radius can be
accommodated by increasing the stellar radius and reducing the
orbital inclination.  In Mazeh et al. (2000),
we undertook a detailed study for of the stellar parameters, 
as well as the respective uncertainties.  These allowed us to 
measure the planetary radius to be $R_{p} = 1.40 \pm 0.17 \ R_{\rm Jup}$.
This value for the radius is consistent with early predictions (Guillot
et al. 1996).  The inflated value of the radius relative to that
of Jupiter is a result of the slower rate of planetary
contraction due to exposure to high stellar insolation soon
after the formation of the planet (Burrows et al. 2000).
We also calculated several derived quantities, in particular
the average density $\rho = 0.31 \pm 0.07 \ {\rm g \, cm^{-3}}$, 
surface gravity $g = 870 \pm 160 \ {\rm cm \, s^{-2}}$,
and escape velocity $v_{e} = 42 \pm 4 \ {\rm km \, s^{-1}}$ of the planet.

\begin{figure}
\plotone{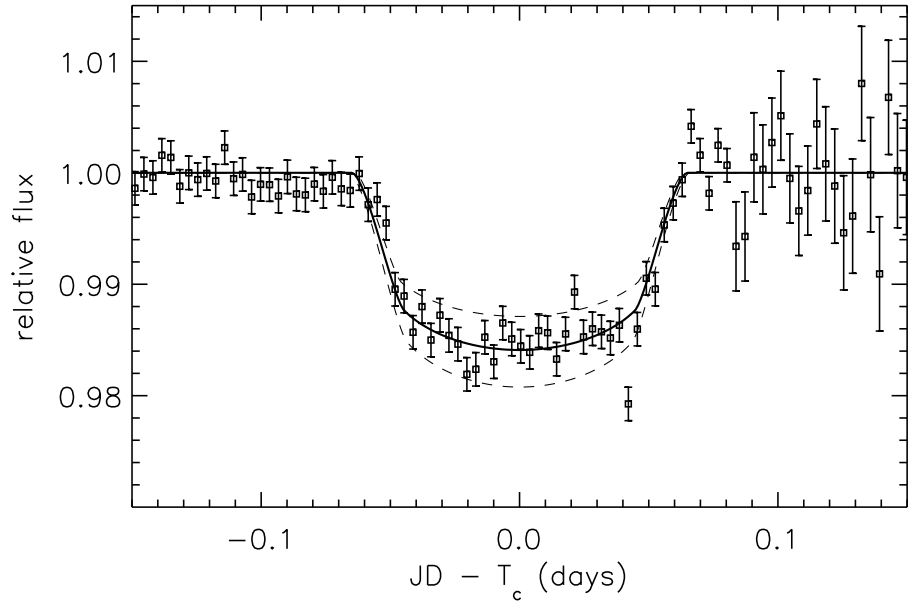}
\caption{STARE observations of the planetary transit of HD~209458,
binned into 5~m averages.
The rms variation at the beginning of the time series
is roughly 1.5 millimag, and this precision
is maintained throughout the duration of the transit.  The increased
scatter at the end of the time series is due to increasing airmass.  
The solid line is the transit
shape that would occur for our best fit model.  The lower and upper dashed
lines are the transit curves that would occur for a planet 
10\% larger and smaller in radius, respectively.  The rapid initial fall 
and final rise of the transit curve correspond
to the times when the planet is crossing 
the edge of the stellar disk, 
and the curvature across the center of the transit
is due to the stellar limb-darkening.  These data are available electronically
from the STARE project website (Brown \& Kolinski 1999).} 
\end{figure}

\section{Follow-Up Observations}
The existence of a transiting planet suggests many fruitful follow-up
observations, some of which may be accomplished within the next
six months:
\begin{itemize}
\item{High cadence, high precision photometry
in other band passes would break the degeneracy
shared between the stellar and planetary parameters, by exploiting
the color-dependent limb-darkening of the star.  Furthermore,
it may be possible to observe color-dependent variations in the
observed planetary radius, since the planet would appear slightly
larger when observed at wavelengths where the atmosphere contains
strong opacity sources (Brown 2000; Burrows et al. 2000).}
\item{If there are other planets in approximately coplanar orbits, 
then the likelihood that they too will generate transits is 
substantially enhanced relative to
that for a randomly oriented system.
A central transit by a Uranus-sized planet at 0.2 AU would yield
a dimming some 6 hours in duration, with a depth of about 1 millimag.}
\item{Reflected light observations such as those for the $\tau$ Bo\"o system 
by Cameron et al. (1999) and Charbonneau et al. (1999) would, if successful, 
yield the planetary albedo directly (Charbonneau \& Noyes 2000).  
Predicted values for the albedo
are highly sensitive to the atmospheric chemistry and condensates 
(Marley et al. 1999; Seager 2000; Seager, Whitney \& Sasselov 2000; 
Sudarsky, Burrows, \& Pinto 2000).}
\item{Observations at wavelengths longer than a few microns may
detect the secondary eclipse (perhaps 4 millimag) 
as the planet passes behind the star.
This would allow the planet's dayside temperature
to be estimated, and hence quantify the net energy deposition in the
planetary atmosphere.
Similarly, it may be feasible to measure the reduction of
the primary eclipse depth in the IR relative to shorter wavelengths
and hence measure the planet's nightside temperature.  If one was able to
determine the dayside and nightside temperatures, then one
would learn if the planetary atmosphere is effective at redistributing
heat due to stellar insolation with a time scale less than the
rotational period.}
\item{If high cadence photometry with a signal-to-noise ratio
of 0.1~millimag can be achieved, it would be possible to detect planetary
rings and/or large rocky satellites (Sartoretti \& Schneider 1999).}
\item{By taking the ratio of high-precision spectra in and out of transit,
it may be possible to see additional absorption features during transit due 
to the absorption of light passing through the limb of the planetary
atmosphere (Brown 2000; Seager \& Sasselov 2000).  The amplitude of the
features may be as large as 0.2\% in the case of the alkali metal lines,
and will be very sensitive to the height of the cloud layer in the
planetary atmosphere.}
\end{itemize}

\acknowledgements
We thank G. Card, C. Chambellan, 
D. Kolinski, A. Lecinski, R. Lull,
T. Russ, and K. Streander
for their assistance in the fabrication, maintenance, and operation of
the STARE photometric camera, and we thank R. Noyes and S. Jha
for many valuable discussions.  D.~ C. is supported 
in part by a Newkirk Fellowship of the High Altitude Observatory.
This work was supported in part by NASA grant W-19560.

\end{document}